\documentstyle[aps,twocolumn,epsfig]{revtex}
\begin{document}
\draft
\tighten
\title{Transverse energy dependence of J/Psi suppression in Au+Au
collisions at RHIC energy}
\author{\bf A. K. Chaudhuri\cite{byline}}
\address{ Variable Energy Cyclotron Centre\\
1/AF,Bidhan Nagar, Calcutta - 700 064\\}
\date{\today}

\maketitle

\begin{abstract}
Prediction  for  transverse  energy  dependence  of  $J/\psi$  to
Drell-Yan ratio in Au+Au collisions at RHIC energy  was  obtained
in  a  model  which  assume  100\% absorption of $J/\psi$ above a
threshold density. The threshold density was obtained by  fitting
the  NA50 data on $J/\psi$ suppression in Pb+Pb collisions at SPS
energy.  At  RHIC  energy,  hard  processes  may  be   important.
Prediction   of   $J/\psi$  suppression  with  and  without  hard
processes were obtained. With hard processes  included,  $J/\psi$'s
are strongly suppressed. \end{abstract}

\pacs{PACS numbers: 25.75.-q, 25.75.Dw}

In  relativistic  heavy  ion  collisions $J/\psi$ suppression has
been recognized as an important tool  to  identify  the  possible
phase transition to quark-gluon plasma. Because of the large mass
of  the  charm  quarks,  $c\bar{c}$ pairs are produced on a short
time scale. Their tight binding also make them  immune  to  final
state interactions. Their evolution probes the state of matter in
the  early  stage  of the collision. Matsui and Satz \cite{ma86}
predicted that in presence of quark-gluon plasma  (QGP),  binding
of $c\bar{c}$ pairs into $J/\psi$ meson will be hindered, leading
to  the  so  called  $J/\psi$ suppression in heavy ion collisions
\cite{ma86}. Over the years  several  groups  have  measured  the
$J/\psi$  yield in heavy ion collisions (for a review of the data
and  the  interpretations  see   \cite{vo99,ge99}).   In   brief,
experimental  data  do  show  suppression.  However this could be
attributed to the conventional nuclear absorption,  also  present
in pA collisions.

The latest data obtained by the NA50 collaboration \cite{na50} on
J/$\psi$ production in Pb+Pb collisions at 158 A GeV is the first
indication  of  anomalous  mechanism  of  charmonium suppression,
which  goes  beyond  the  conventional  suppression  in   nuclear
environment.  The  ratio  of  J/$\psi$ yield to that of Drell-Yan
pairs decreases faster with $E_T$ in the most central  collisions
than  in  the  less  central ones. It has been suggested that the
resulting pattern can be understood in a  deconfinement  scenario
in  terms  of  successive  melting  of  charmonium  bound  states
\cite{na50}.

In  a recent paper Blaizot et al \cite{bl00} showed that the data
can be understood as an effect of transverse energy  fluctuations
in   central   heavy   ion   collisions.   Introducing  a  factor
$\varepsilon=E_T/E_T(b)$ and assuming  that  the  suppression  is
100\%  above  a  threshold density (a parameter in the model) and
smearing  the  threshold  density  (at  the  expense  of  another
parameter)  best  fit  to  the  data  was obtained. Capella et al
\cite{ca00} analysed the data  in  the  comover  approach.  There
also,  the  comover  density  has  to  be  modified by the factor
$\varepsilon$. Introduction of this  adhoc  factor  $\varepsilon$
can be justified in a model based on excited nucleons represented
by strings \cite{hu00}.

At a fixed impact parameter, the transverse energy as well as the
number of NN collisions fluctuate. The Fluctuations in the number
of  NN collisions were not taken into account in the calculations
of Blaizot et al \cite{bl00} or in the calculations of Capella et
al \cite{ca00}. We have analysed  \cite{ch01}  the  NA50  data  ,
extending  the  model of Blaizot et al \cite{bl00} to include the
fluctuations  in  number  of  NN  collisions  at   fixed   impact
parameter.  The  $E_T$ distribution was obtained in the geometric
model, which includes these fluctuations. It was shown that  with
a  single  parameter,  the  threshold  density,  above  which the
$J/\psi$ suppression is assumed to be 100\%, good fit to the data
can be obtained. In the present paper, we have applied the model
to predict the $E_T$ dependence of $J/\psi$ to Drell-Yan ratio at
RHIC energy. With RHIC being operational, it is  hoped  that  the
prediction  will  help  to plan experiments to detect QGP. In the
following, we will present in brief the model. The details can be
found in \cite{ch01}.

In   \cite{ch01}  $E_T$  distribution  for  Pb+Pb  collision  was
obtained in the Geometric model \cite{ch90,ch93}. In this  model,
$E_T$  distribution of AA collisions is written in terms of $E_T$
distribution in NN collisions. One  also  assume  that  the  Gama
distribution  with  parameters  $\alpha$ and $\beta$ describe the
$E_T$-distribution in NN collisions. $E_T$ distribution for Pb+Pb
could  be  fitted  well   with   $\alpha=3.46   \pm   0.19$   and
$\beta=0.379\pm0.021$.  In  fig.1a,  the  experimental data along
with the fit are shown. Parametric values of $\alpha$ and $\beta$
indicate that average $E_T$ produced in individual NN  collisions
is $\beta/\alpha \sim .1 GeV $. This is to be contrasted with the
average  $E_T  \sim  1  GeV  $  produced  in  other AA collisions
\cite{ch93}.

As  mentioned  in  the  beginning,  we  have assumed that above a
threshold density $n_c$,  the  charmonium  suppression  is  100\%
effective  \cite{bl00}.  Charmonium  production  cross-section at
impact parameter ${\bf b}$ is written as,

\begin{equation}
d^2\sigma^{J/\psi}/d^2b     =     \sigma^{J/\psi}    \int    d^2s
T^{eff}_A({\bf s}) T^{eff}_B({\bf s}-{\bf b}) S({\bf b},{\bf  s})
\label{6b}
\end{equation}

\noindent   where   $T_{A,B}^{eff}$   is  the  effective  nuclear
thickness function,

\begin{equation} \label{6c}
T^{eff}      ({\bf s})=\int_{-\infty}^\infty dz \rho({\bf s},z)
exp(-\sigma_{abs}\int_z^\infty dz^\prime \rho({\bf s},z^\prime))
\end{equation}

\noindent  with  $\sigma_{abs}$ as the cross-section for $J/\psi$
absorption by nucleons. The exponential  factor  is  the  nuclear
absorption   survival   probability,   the  probability  for  the
$c\bar{c}$ pair to avoid nuclear absorption and form a  $J/\psi$.
$S({\bf  b},{\bf  s})$  is the anomalous part of the suppression.
Blaizot et al \cite{bl00} assumed that  $J/\psi$  suppression  is
100\% effective above a threshold density ($n_c$), a parameter in
the model. Accordingly the anomalous suppression part was written
as,

\begin{equation}
S({\bf b},{\bf s})= \Theta(n_c - \varepsilon n_p({\bf b},{\bf s}))
\label{7}
\end{equation}

\noindent  where  $n_p$ is the density of participant nucleons in
the impact parameter space,

\begin{equation}
n_p({\bf      b},{\bf     s})=     T_A({\bf
s})[1-e^{-\sigma_{NN}   T_B({\bf    b}-{\bf    s})}]    +    [T_A
\leftrightarrow T_B] \label{8} \end{equation}

\noindent   and $\varepsilon = E_T/E_T(b)
=E_T/n\beta/\alpha$ is  the
modification  factor  which  takes  into  account  the transverse
energy fluctuations at fixed impact parameter  \cite{bl00}.  This
modification  makes  sense  only  when  $n_p$  is  assumed  to be
proportional to  the  energy  density.  Implicitly  it  was  also
assumed  that  the  $E_T$ fluctuations are strongly correlated in
different rapidity gaps. The assumption  was  essential  as  NA50
collaboration measured $E_T$ in the 1.1-2.3 pseudorapidity window
while  the  $J/\psi$'s  were  measured  in  the  rapidity  window
$2.82<y<3.92$  \cite{na50}.  Strong  correlation  between   $E_T$
fluctuations  in  different  rapidity windows is explained in the
Geometric model \cite{ch01}.

We  calculate the $J/\psi$ production as a function of transverse
energy, at an impact parameter ${\bf b}$ as,

\begin{equation}\label{1a}
d \sigma  ^{J/\psi}/dE_T=\sum  _{n=1}^\infty
P_n(b,E_T)  d^2\sigma^{J/\psi}/d^2b
\end{equation}

\noindent  where  $P_n(b,E_T)$ is the probability to obtain $E_T$
in $n$ NN collisions,  expression  for  which  can  be  found  in
\cite{ch01}.

The  Drell-Yan  production  was  calculated  similarly, replacing
charmonium  cross  section  in  eq.\ref{1a}  by   the   Drell-Yan
cross-section,

\begin{equation}
d^2\sigma^{DY}/d^2b   =   \sigma^{DY}   \int   d^2s
T_A({\bf s}) T_B({\bf s}-{\bf b}) \label{6a}
\end{equation}

In fig.1b, we have compared the theoretical charmonium production
cross-section  with  NA50  experimental  data.  The normalization
factor $\sigma^{J/\psi}/\sigma^{DY}$ was taken to  be  53.5.  The
solid  curve is obtained with $\sigma_{abs}$=6.4 mb, and $n_c=3.8
fm^{-2}$. Very good description of the data from 40 GeV onward is
obtained. The model reproduced the 2nd drop around 100 GeV,  (the
knee  of  the  $E_T$  distribution).  It may be noted that if the
fluctuations in the  NN  collisions  were  neglected,  equivalent
description  is  obtained with threshold density $n_c=3.75 fm^2$,
with smearing of the $\Theta$ function at the expense of  another
parameter. It is evident that in this model, the smearing is done
by  fluctuating  NN  collisions. Theoretical calculations predict
more suppressions below 40 GeV, a feature evident in other models
also. It is possible to fit the entire $E_T$ range, reducing  the
$J/\psi$-nucleon    absorption    cross-scetion.    Recent   data
\cite{le00} on the $J/\psi$  cross  section  in  $pA$  collisions
point  to a smaller value of $\sigma_{abs}$ $\sim$ 4 to 5 mb. The
dashed line in fig.1b,  corresponds  to  $\sigma_{abs}$=4  mb  and
$n_c=3.42   fm^{-2}$.   However,   it   may   be  mentioned  that
$\sigma_{abs}$=4 mb does not allow a good fit to the $pA$ and S-U
data \cite{kh97}.

\begin{figure}[h]
\centerline{\psfig{figure=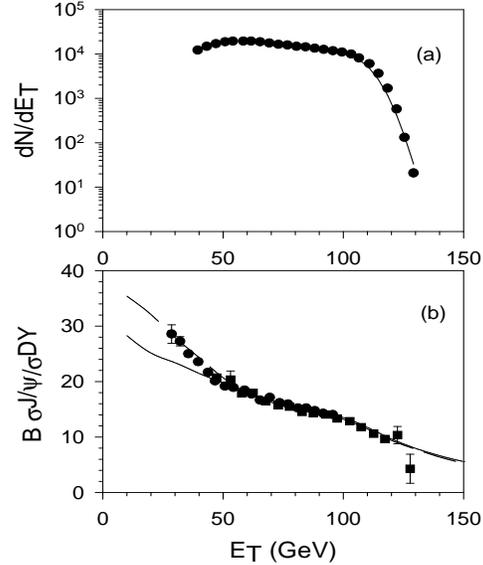,height=10cm,width=8cm}}
\vspace{-1cm}
\caption{(a) Transverse energy distribution in Pb+Pb collisions,
(b) $J/\psi$ to Drell Yan ratio in  Pb+Pb  collisions  as  a
function of transverse energy.}
\end{figure}

Present model can be used to predict $E_T$ dependence of $J/\psi$
to  Drell-Yan  ratio  at  RHIC  energy.  Recent PHOBOS experiment
\cite{phobos}  showed  that   for   central   collisions,   total
multiplicity  is  larger  by  70\% at RHIC than at SPS. We assume
that $E_T$ is correspondingly increased \cite{bl01}. Accordingly,
we rescale the $E_T$ distribution for Pb+Pb collisions and assume
that it represent the experimental $E_T$ distribution  for  Au+Au
collisions  at  RHIC  (small mass difference between Au and Pb is
neglected). At RHIC energy the so-called hard component which  is
proportional   to  number  of  binary  collisions  appear.  Model
dependent calculations indicate that  the  hard  component  grows
from 22\% to 37\% as the energy changes from $\sqrt{s}$=56 GeV to
130  GeV  \cite{kh01}.  However  we  choose  to  ignore  the hard
component in $E_T$-distribution.  The  multiplicity  distribution
obtained  by  the  PHOBOS  collaboration,  in  the rapidity range
$3<\mid \eta \mid <4.5$ could be fitted well with or without this
hard component. Indeed, it  appears  that  the  data  are  better
fitted   without   the   hard   component   \cite{kh01}.   Global
distribution e.g. multiplicity or transverse energy distributions
are not sensitive  to  the  hard  component.  In  fig.2a,  filled
circles represent the "experimental" $E_T$ distribution for Au+Au
collisions at RHIC, obtained by scaling the $E_T$ distribution in
Pb+Pb  collisions  at  SPS.  The  solid  line  is  a  fit  to the
"experimental"  $E_T$-distribution  in   the   geometric   model,
obtained  with  $\alpha$=3.09  and $\beta$=0.495. Nucleon-nucleon
inelastic cross section ($\sigma_{NN}$) was assumed to be  41  mb
at  RHIC,  instead  of 32 mb at SPS \cite{bl01}. Fitted values of
$\alpha$ and $\beta$ are interesting. With these  values  average
$E_T$   produced   in   individual   NN  collisions  at  RHIC  is
$\beta/\alpha$=0.16 GeV. This is to be compared  with  the  value
0.1  GeV  for  Pb+Pb collisions at SPS. Average $E_T$ produced in
individual NN collisions at RHIC is increased by  60\%,  compared
to  SPS  energy.  The  apparent  inconsistency  is resolved if we
remember that $\sigma_{NN}$ is increased by 30\% from SPS to RHIC
energy.

\begin{figure}[h]
\centerline{\psfig{figure=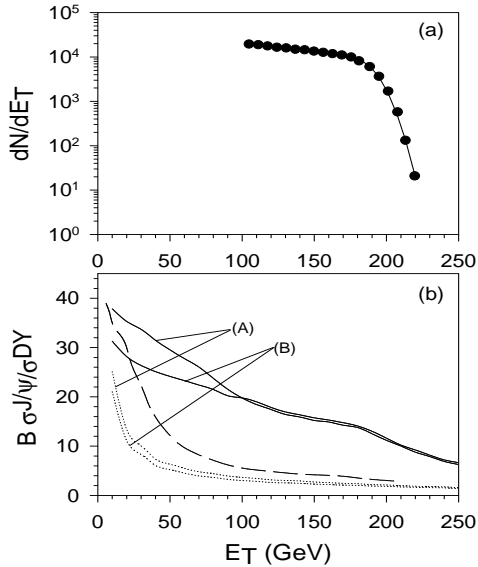,height=10cm,width=8cm}}
\vspace{-1cm}
\caption{  (a)Expected  $E_T$ distribution in Au+Au collisions at
RHIC energy. The solid line is a fit in the geometric model  (see
text),  (b)  Expected  $E_T$  dependence of $J/\psi$ to Drell-Yan
ratio in Au+Au collisions at  RHIC  energy.  The  dotted  (solid)
lines  are  the  predictions  with  (without) the hard scattering
component, for two sets of parameters. The The  dashed  curve  is
the prediction of Blaizot et al \protect\cite{bl01}}
\end{figure}

While  the $E_T$ distribution at RHIC may not be sensitive to the
hard scattering component, the $J/\psi$ suppression will  be.  In
the  model  $J/\psi$ suppression is 100\% if the $n_p E_T/E_T(b)$
exceeds the threshold density $n_c$. Without the hard  component,
transverse  density  in  Au+Au  collisions will be nearly same as
that in Pb+Pb collisions at SPS. At RHIC, $E_T(b)$ is  increased,
individual  NN  collisions  produces  more  $E_T$. Then anomalous
suppression will set at larger  $E_T$  (threshold  density  being
same).  However,  with the hard component, the transverse density
will be modified. For $f$ fraction of hard scattering  transverse
density can be written as \cite{bl01},

\begin{equation}
n_p({\bf  b,s})  \rightarrow  (1-f) n_p({\bf b,s}) + 2 f
n^{hard}_p({\bf b,s})
\end{equation}

\noindent  with  $n^{hard}_p({\bf  b,s})=\sigma_{NN}  T_A({\bf s})
T_B({\bf b-s})$.  With  hard  component,  transverse  density  is
increased,  as  a  result,  anomalous  suppression will set in at
lower $E_T$.

In  fig.2b,  we  have  presented  the $J/\psi$ to Drell-Yan ratio
obtained in the present model for Au+Au collision at RHIC energy.
We have presented the results for the two sets of parameters;(A)
$\sigma_{abs}$=6.4    mb,    $n_c$=3.8    $fm^{-2}$    and   (B)
$\sigma_{abs}$=4 mb, $n_c$=3.44 $fm^{-2}$.
The  solid  lines  are  the  prediction  for $J/\psi$ suppression
neglecting the hard component in the transverse density, for  the
two sets of parameters (A) and (B) respectively. The dotted lines
are  the  prediction  including  the hard component. We find that
without the hard  component,  $J/\psi$  suppression  at  RHIC  is
similar  to  that  obtained  at  SPS  energy.  The 2nd drop which
occurred at 100 GeV in Pb+Pb collisions at SPS energy,  now  sets
in  around  180  GeV (knee of the $E_T$ distribution being around
that  energy).  As  mentioned  earlier,  at  RHIC,  $E_T(b)$   is
increased,  and without the hard component the transverse density
is nearly same as it was in Pb+PB collisions  at  SPS.  Anomalous
suppression  then  occurs  at  larger  $E_T$. As it was for Pb+Pb
collisions at SPS, the two sets give nearly same suppression  for
$E_T$  beyond  100  GeV.  It is also an indication that anomalous
suppression occur at larger $E_T$.

If  the  transverse  density  is  modified  to  include  the hard
component, $J/\psi$'s are strongly suppressed. In fig.2b, the dotted
lines presents the results obtained with 37\% \cite{kh01} hard
component.  Nearly  same suppression is obtained for set A and B.
At knee, suppression  is  6  times  greater  than  corresponding
suppression  without the hard scattering component. Very strong suppression wash out
the 2nd drop, which  was  clearly  visible  at  SPS,  or  in  the
prediction  without  the  hard component. With hard component, as
mentioned earlier, transverse density is increased and  anomalous
suppression  sets  in  earlier.  Blaizot  et  al \cite{bl01} also
predicted the $E_T$ dependence of the  $J/\psi$  suppression.  In
fig.2b,  their  result  is  shown (the dashed line). As mentioned
earlier, fluctuations in number of NN collisions at fixed  impact
parameter  was neglected. Also the model for $E_T$ production was
not microscopic. However, at large $E_T$, our prediction  agrees  closely  with
theirs.

To  summarize,  prediction  for  $J/\psi$  suppression  in  Au+Au
collisions at RHIC energy was obtained in a model  which  assume
100\%   absorption   of   $J/\psi$  above  a  threshold  density.
Transverse energy fluctuations as well  as  fluctuations  in  the
number of NN collisions at fixed impact parameter were taken into
account.  The threshold density was obtained from the analysis of
NA50 data on $J/\psi$ suppression  in  Pb+Pb  collisions  at  SPS
energy.  At  RHIC energy hard processes are important. Prediction
of $J/\psi$ suppression, with  and  without  the  hard  processes
differ  considerably.  Without hard processes, predicted $J/\psi$
suppression at RHIC energy is similar to  that  obtained  at  SPS
energy.  The  2nd  drop  which occur at $E_T \sim$ 100 GeV at SPS
energy moves upward to  180  GeV.  Inclusion  of  hard  processes
modifies  the transverse density resulting in considerable larger
suppression. Very large suppression  washes  out  the  2nd  drop,
which  was visible at SPS energy or in the prediction without the
hard processes.

 \end{document}